\documentclass[10pt]{article}
\usepackage{epsfig}
\usepackage{lineno}
\usepackage{cite}

\newcommand{\be}{\begin{equation}}
\newcommand{\ee}{\end{equation}}
\newcommand{\bea}{\begin{eqnarray}}
\newcommand{\eea}{\end{eqnarray}}

\begin{document}

\title{ \vspace{1cm} Strength of the  E$_{R}$ = 127 keV, $^{26}$Al(p,$\gamma$)$^{27}$Si resonance}
\author{A. Parikh$^{1,2,} \footnote{anuj.r.parikh@upc.edu}$ , J. Jos\'e$^{1,2}$, A. Karakas$^{3}$, C. Ruiz$^{4}$, K. Wimmer$^{5}$\\
\\
\small $^1$Departament de F\'isica i Enginyeria Nuclear, \\ \small Universitat
Polit\`ecnica de Catalunya, EUETIB, E-08036 Barcelona, Spain\\
\small $^2$Institut d'Estudis Espacials de Catalunya (IEEC), E-08034 Barcelona, Spain\\
\small $^3$Research School of Astronomy and Astrophysics, Australian National University, Canberra, ACT 2611, Australia\\
\small $^4$TRIUMF, Vancouver, British Columbia V6T 2A3, Canada\\
\small $^5$Department of Physics, Central Michigan University, Mount Pleasant, Michigan 48859, USA\\
}
\maketitle

\begin{abstract} We examine the impact of the strength of the E$_{R}$ = 127 keV, $^{26}$Al($p,\gamma$)$^{27}$Si resonance on $^{26}$Al production in classical nova explosions and asymptotic giant branch (AGB) stars.  Thermonuclear $^{26}$Al($p,\gamma$)$^{27}$Si reaction rates are determined using different assumed strengths for this resonance and representative stellar model calculations of these astrophysical environments are performed using these different rates.  Predicted $^{26}$Al yields in our models are not sensitive to differences in rates determined using zero and a commonly stated upper limit corresponding to $\omega\gamma_{UL} = 0.0042$ $\mu$eV for this resonance strength.  Yields of $^{26}$Al decrease by 6\% and, more significantly, up to 30\%, when a strength of $24 \times \omega\gamma_{UL} = 0.1$ $\mu$eV is assumed in the adopted nova and AGB star models, respectively.  Given that the value of $\omega\gamma_{UL}$ was deduced from a single, background-dominated $^{26}$Al($^{3}$He,d)$^{27}$Si experiment where only upper limits on differential cross sections were determined, we encourage new experiments to confirm the strength of the 127 keV resonance.      
\end{abstract}

{\bf PACS number(s):} 26.30.-k, 26.20.Np, 27.30.+t

\twocolumn

The origin of the observed Galactic radioisotope $^{26}$Al is still unresolved.  Over thirty years have passed since the first identification~\cite{Mah82} in the Galactic interstellar medium of the 1.809-MeV $\beta$-delayed $\gamma$-ray
line from the decay of the ground state of $^{26}$Al (t$_{1/2}$ = 7.17 $\times 10^{5}$ y).  Since then, increasingly sophisticated observational studies have produced all-sky maps of the 1.809 MeV emission (showing that $^{26}$Al is mostly confined to the Galactic disk)\cite{Plu01}, demonstrated that $^{26}$Al co-rotates with the Galaxy (supporting its Galaxy-wide origin)\cite{Die06}, and used $^{26}$Al as a tracer to examine the kinematics of massive star and supernova ejecta\cite{Kre13}, among other achievements.  The stellar production of $^{26}$Al has also been inferred through measured excesses of its daughter $^{26}$Mg in inclusions and presolar dust grains within primitive meteorites\cite{Hop94,Mac95,Zin07,Mac14}.  Nonetheless, despite extensive theoretical studies of nucleosynthesis in proposed astrophysical environments\cite{Pra96, Die98} such as asymptotic giant branch (AGB) stars\cite{Mow00, Kar03,Izz07}, classical nova explosions\cite{Coc95,Jos97,Jos99} and massive stars\cite{Lim06, Woo07, Ili11}, accounting for the present-day Galactic $^{26}$Al abundance of $2-3$ M$_{\odot}$\cite{Die06,Die10} has proved elusive.

In hydrogen-burning environments, an accurate thermonuclear rate of the $^{26}$Al($p,\gamma$)$^{27}$Si destruction reaction at the relevant stellar temperatures is clearly needed for reliable model predictions of $^{26}$Al production.  For example, according to current models, winds from AGB stars eject $^{26}$Al produced at temperatures of $\approx50 -100$ MK, while in classical novae, $^{26}$Al is produced in explosions that involve an oxygen-neon white dwarf and achieve peak temperatures of T$_{peak}$  $\approx0.2-0.4$ GK.  To determine the $^{26}$Al($p,\gamma$) rate in these environments, one therefore requires resonance energies $E_{R}$ (which enter exponentially in the rate) and ($p,\gamma$) resonance strengths $\omega\gamma$ (which enter linearly in the rate) for $^{27}$Si states between the $^{26}$Al+p energy threshold (S$_{p}$ =  7463.25(16) keV \cite{AME12}) and $\approx$200 and $\approx$500 keV above this threshold for AGB stars and novae, respectively.   [Note that throughout this manuscript we discuss exclusively the ($p,\gamma$) reaction on the 5$^{+}$ ground state of $^{26}$Al rather than on the 0$^{+}$ isomeric state at E$_x$ = 228 keV (t$_{1/2}$ = 6.3 s).]

The principal uncertainties in the $^{26}$Al($p,\gamma$)$^{27}$Si rate at temperatures relevant to $^{26}$Al production in AGB stars and novae arise from the unmeasured strengths of the resonances at E$_{R}$ = 68 and 127 keV\cite{Ili10}.  Tentative observations of additional states\cite{Wan89,Lot11}, which would correspond to E$_{R}$ = 30 and 94 keV, should also be confirmed, although we note that two relatively non-selective, recent indirect studies did not observe the latter level\cite{Lot11,Par11}.  While it may play a role in AGB stars, the uncertainty in the rate due to the strengths of resonances at 30, 68 and 94 keV is not expected to significantly affect $^{26}$Al production in novae\cite{Coc95,Jos99}.  Therefore, in the present work we focus on the impact of variations in the strength of the 127 keV resonance on $^{26}$Al production in models of AGB stars and classical novae.  Obviously any sensitivity of $^{26}$Al production in these environments to reasonable adopted strengths for this one resonance would only be exacerbated through consideration of additional resonances.  

Most detailed nova and AGB star models that have examined the production of $^{26}$Al\cite{Coc95,Jos97,Jos99,Rui06,Izz07,Kar10, Ben13} have used $^{26}$Al($p,\gamma$) rates that incorporate a result from Vogelaar et al.~(1996)~\cite{Vog96} for the strength of the 127 keV resonance. Indeed, studies have estimated that novae may contribute up to $\approx30\%$ of the Galactic $^{26}$Al abundance using such a rate\cite{Jos97,Ben13}.  Vogelaar et al. measured differential cross sections for $^{27}$Si states above the $^{26}$Al+p energy threshold populated through the $^{26}$Al($^{3}$He,d)$^{27}$Si proton-transfer reaction.  Assuming purely single-particle transfer, ($p,\gamma$) resonance strengths may be estimated from proton spectroscopic factors C$^2$S extracted from such an experiment.  For the state at E$_{x}$ = 7590 keV (E$_{R}$ = 127 keV) solely upper limits for differential cross sections were determined, and at only three of the nine angles at which measurements were made.  These limited results were largely due to the nature of the target employed, which was dominated by $^{27}$Al ($^{26}$Al/$^{27}$Al = 6.3\%).  Because of this, the measured deuteron spectra were dominated by products from the competing $^{27}$Al($^{3}$He,d) reaction.   Direct reaction calculations assuming $l=0$ transfer were then used with the upper limits on the differential cross sections to give their stated upper limit of C$^{2}$S$_{UL}$ = 0.002 for this $9/2^{+}$\cite{Lot11,Par11} state.  This would correspond to a strength of $\omega\gamma_{UL}$ = 0.0042 $\mu$eV for the E$_{R}$ = 127 keV resonance under the reasonable assumption that the proton partial width for this threshold state is much less that the $\gamma$-ray partial width.

\begin{figure}[h]
\begin{center}
\includegraphics[scale=0.6]{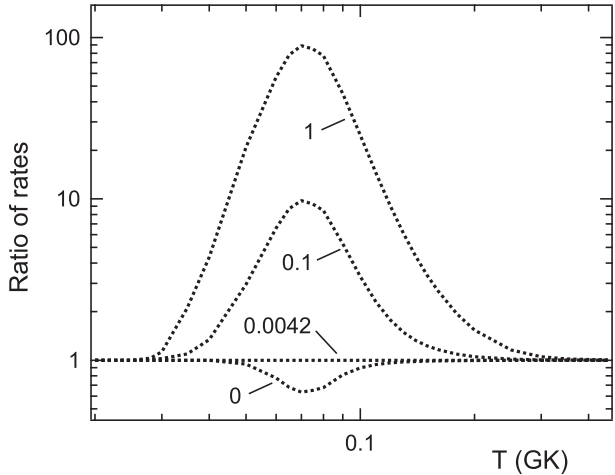}
\caption{Thermonuclear rates of the $^{26}$Al(p,$\gamma$)$^{27}$Si reaction calculated using different assumed strengths (in $\mu$eV) for the E$_{R}$ = 127 keV resonance.  Rates are shown relative to the rate calculated using a strength of 0.0042 $\mu$eV.  All other parameters for these rate calculations were adopted from Ref.\cite{Ili10}.}
\label{fig1}
\end{center}
\end{figure}

This upper limit on the spectroscopic factor may be questionable for several reasons.  The dearth of angles at which differential cross section upper limits were extracted for this state makes the theoretical fit highly dependent on the reliability of the limit at the lowest angle ($\theta_{c.m.}\approx5^{\circ}$, see Fig. 6a in Ref.\cite{Vog96}).  If, instead, their calculation were scaled to the upper limit at the highest angle at which a limit was extracted from the background-dominated spectra ($\theta_{c.m.}\approx14^{\circ}$), the C$^{2}$S value would increase by a factor of $\approx20$.   
Furthermore, we have repeated the $l=0$ theoretical calculation for this state using the direct reaction code FRESCO\cite{thompson88} and we find a C$^{2}$S value up to $\approx5$ times larger than that of Vogelaar et al. when reasonable sets of optical model parameters are adopted\cite{Vog96,perey,pang}.  Finally, for the $9/2^{+}$\cite{Lot11,Par11}, 7739 keV $^{27}$Si state, Vogelaar et al. determine a strength for an $l=0$ transition (via an extracted C$^{2}$S) that differs from the directly-measured value\cite{Buc84} by a factor of $\approx5$.  For this state, $^{26}$Al($^{3}$He,d) cross sections were measured at seven angles.  While the discrepancy may be due to, for example, population of this state through a mixed transition, an erroneous spectroscopic factor due to issues with the measured differential cross sections or the theoretical calculations cannot be ruled out.  With regard to the thermonuclear rate of the $^{26}$Al($p,\gamma$)$^{27}$Si reaction, a spectroscopic factor of zero for the 127 keV state leads to a rate up to 1.6 times lower than that determined using $\omega\gamma_{UL}$ over T = 0.05 -- 0.11 GK.  On the other hand, a strength of $\omega\gamma = 24 \times \omega\gamma_{UL}$ =  0.1 $\mu$eV (or equivalently, C$^{2}$S = 24 $\times$ C$^{2}$S$_{UL}$) has a dramatic effect on the rate over T = 0.04 -- 0.2 GK, leading to enhancements by as much as a factor of 10 relative to the rate using $\omega\gamma_{UL}$.  Such an enhanced rate would not be unreasonable given the above discussion.  These reaction rates are shown in Fig. \ref{fig1}.  

To assess the sensitivity of model predictions of $^{26}$Al yields to the strength of the E$_{R}$ = 127 keV, $^{26}$Al($p,\gamma$)$^{27}$Si resonance, we have performed new sets of representative hydrodynamic nova models and stellar nucleosynthesis calculations for AGB stars.  We have used these models together with $^{26}$Al($p,\gamma$)$^{27}$Si rates calculated assuming E$_{R}$ = 127 keV resonance strengths of 0, 0.0042 $\mu$eV, and 0.1 $\mu$eV, as discussed above.  To fully explore the impact of this strength, we have also used a rate determined using a strength of 1 $\mu$eV, although we note that the corresponding C$^{2}$S value of 0.5 seems incompatible with the data of Ref.\cite{Vog96}.  For the nova simulations, a 1.25 M$_\odot$ oxygen-neon white dwarf was evolved from the
accretion stage to the explosion, expansion and ejection stages.  Four models,
identical except for the adopted prescription of the $^{26}$Al($p,\gamma$) rate, have been computed
with the spherically symmetric, implicit, Lagrangian code SHIVA, extensively 
used in the modeling of stellar explosions such as classical novae and type I X-ray bursts \cite{JH98}.
The solar-like accreted material was pre-mixed with material from the outer layers of the white dwarf at a level of
50\% to mimic mixing at the core-envelope interface\cite{Cas11}.  Typical values for the
initial white dwarf luminosity ($10^{-2}$ L$_\odot$) and the mass-accretion rate  
($2 \times 10^{-10}$ M$_\odot$ yr$^{-1}$)  have been adopted, resulting in explosions with T$_{peak}$ = 0.25 GK.  Nucleosynthesis in AGB stars was examined using models of 6 M$_{\odot}$ and 8 M$_{\odot}$ stars, with metallicities of Z = 0.004 and 0.014, respectively\cite{Kar10,Kar14}.  These models were chosen because the temperature at the base of the convective envelope reaches $\approx$0.1 GK during the thermally-pulsing AGB phase, which makes them ideal sites for testing the impact of reaction rates related to the production of $^{26}$Al.  Abundances in the AGB star models were determined with a post-processing algorithm\cite{Kar10} that incorporates time-dependent diffusive mixing for all convective zones\cite{Can93}.  

Models using $^{26}$Al($p,\gamma$)$^{27}$Si rates determined with $\omega\gamma$ = 0 and $\omega\gamma_{UL}$ = 0.0042 $\mu$eV agreed to better than 3\% in the amount of $^{26}$Al produced, for both the nova and AGB star simulations.   Yields of $^{26}$Al decreased by 6\% and 40\% when the reaction rates calculated with strengths of 0.1 and 1 $\mu$eV were used in the nova models, relative to the $^{26}$Al yield determined using $\omega\gamma_{UL}$.  The impact of the enhanced rates in the AGB star models is rather more striking: for the 8 M$_{\odot}$ model, $^{26}$Al yields decreased by 10\% and a factor of 2 when the rates with strengths of 0.1 and 1 $\mu$eV were employed; for the 6 M$_{\odot}$ model, $^{26}$Al yields decreased by 30\% and a factor of 6 when the rates with strengths of 0.1 and 1 $\mu$eV were used, all relative to the $^{26}$Al yield determined using $\omega\gamma_{UL}$.  We also note that in a study of the impact of reaction rate variations on $^{26}$Al production in massive stars\cite{Ili11}, an enhancement of the $^{26}$Al($p,\gamma$)$^{27}$Si rate by a factor of 10 during core hydrogen burning reduced the predicted $^{26}$Al yield by a factor of 1.8.  As shown in Fig. \ref{fig1}, this level of enhancement of the rate at the relevant temperatures (T $\approx 0.04 - 0.08$ GK\cite{Lim06,Ili11}) would follow from a strength of 0.1 $\mu$eV for the 127 keV resonance. 

Given the impact on model predictions of $^{26}$Al production, we encourage experimental efforts to measure the strength of the E$_{R} = 127$ keV, $^{26}$Al($p,\gamma$)$^{27}$Si resonance.  A new $^{26}$Al($^{3}$He,d)$^{27}$Si measurement with an improved target would be helpful to both confirm the results of Vogelaar et al.\cite{Vog96} for the 127 keV resonance and to help estimate the unknown strengths of the lower energy resonances.  Sufficiently stringent upper limits from direct measurements would also be welcome.


\begin{thebibliography}{99}
\itemsep -2pt 

\bibitem{Mah82}W. A. Mahoney, J. C. Ling, A. S. Jacobson, and R. E. Lingenfelter, Astrophys. J. 262, 742 (1982).
\bibitem{Plu01}S. Pl\"uschke, et al., in \emph{Exploring the Gamma-ray universe}, eds. B. Battrick, A. Gimenez, V. Reglero, and C. Winkler, ESA SP--459, 55 (2001).
\bibitem{Die06}R. Diehl, et al., Nature 439, 45 (2006).
\bibitem{Kre13}K. Kretschmer, R. Diehl, M. Krause, A. Burkert, K. Fierlinger, O. Gerhard, J. Greiner, and W. Wang, Astron. Astrophys. 559, A99 (2013). 
\bibitem{Mac95} G.J. MacPherson, A.M. Davis, and E.K. Zinner, Meteoritics 30, 365 (1995).
\bibitem{Mac14} G.J. MacPherson, A.M. Davis, and E.K. Zinner, 45th Lun. Plan. Sci. Conf., LPI 1777, 2134 (2014).
\bibitem{Hop94} P. Hoppe, S. Amari, E. Zinner, T. Ireland, T., and R.S. Lewis, Astrophys. J. 430, 870 (1994).
\bibitem{Zin07} E. Zinner, Presolar grains, in \emph{Meteorites, Planets, and Comets}, eds. A.M. Davis, H.D. Holland, and K.K. Turekian, Treatise on Geochemistry, vol.1.02, p.1 (Elsevier, Oxford, 2007).
\bibitem{Pra96}N. Prantzos, and R. Diehl, Phys. Rep. 267, 1 (1996).
\bibitem{Die98}R. Diehl, and F.X. Timmes, PASP 110, 637 (1998).
\bibitem{Mow00}N. Mowlavi, and G. Meynet, Astron. Astrophys. 361, 959 (2000).
\bibitem{Kar03}A. Karakas, and J.C. Lattanzio, PASA 20, 279 (2003).
\bibitem{Izz07}R.G. Izzard, M. Lugaro, A.I. Karakas, C. Iliadis, and M. van Raai, Astron. Astrophys. 466, 641 (2007).
\bibitem{Coc95}A. Coc, R. Mochkovitch, Y. Oberto, J.-P. Thibaud, and E. Vangioni-Flam, Astron. Astrophys. 299, 479 (1995).
\bibitem{Jos97} J. Jos\'e, M. Hernanz, and A. Coc, Astrophys. J. 479, L55 (1997).
\bibitem{Jos99} J. Jos\'e, A. Coc, and M. Hernanz, Astrophys. J. 520, 347 (1999).
\bibitem{Lim06} M. Limongi, and A. Chieffi, Astrophys. J. 647, 483 (2006).
\bibitem{Woo07} S.E. Woosley, and A. Heger, Phys. Rep. 442, 269 (2007).
\bibitem{Ili11}C. Iliadis, A. Champagne, A. Chieffi, and M. Limongi, Astrophys. J. Suppl. 193, 16 (2011).
\bibitem{Die10}R. Diehl, M.G. Lang, P. Martin, H. Ohlendorf, Th. Preibisch, R. Voss, P. Jean, J.-P. Roques, P. von Ballmoos, and W. Wang, Astron. Astrophys. 522, A51 (2010). 
\bibitem{AME12} M. Wang, G. Audi, A.H. Wapstra, F.G. Kondev, M. MacCormick, X. Xu, and B. Pfeiffer, Chinese Phys. C36, 1603 (2012).
\bibitem{Ili10}C. Iliadis, R. Longland, A.E. Champagne, A. Coc, and R. Fitzgerald, Nucl. Phys. A841, 31 (2010).
\bibitem{Wan89}T. F. Wang, A. E. Champagne, J. D. Hadden, P. V. Magnus, M. S. Smith, A. J. Howard, and P. D. Parker, Nucl. Phys. A499, 546 (1989).
\bibitem{Lot11}G. Lotay, P.J. Woods, D. Seweryniak, M.P. Carpenter, H.M. David, R.V.F. Janssens, and S. Zhu, Phys. Rev. C 84, 035802 (2011).
\bibitem{Par11}A. Parikh, et al., Phys. Rev. C 84, 065808 (2011).
\bibitem{Rui06}C. Ruiz, et al., Phys. Rev. Lett. 96, 252501 (2006).
\bibitem{Ben13}M.B. Bennett, et al., Phys. Rev. Lett. 111, 232503 (2013).
\bibitem{Kar10}A. Karakas, MNRAS 403, 1413 (2010).
\bibitem{Vog96}R.B. Vogelaar, L.W. Mitchell, R.W. Kavanagh, A.E. Champagne, P.V. Magnus, M.S. Smith, A.J. Howard, P.D. Parker, and H.A. O'Brien, Phys. Rev. C 53, 1945 (1996).
\bibitem{thompson88}I. J. Thompson, Comp. Phys. Rep. 7, 167 (1988).
\bibitem{perey}C. M. Perey, and F. G. Perey, At. Data Nucl. Data Tables 17, 1 (1976).
\bibitem{pang}D.Y. Pang, P. Roussel-Chomaz, H. Savajols, R.L. Varner, and R. Wolski, Phys. Rev. C 79, 024615 (2009).
\bibitem{Buc84} L. Buchmann, M. Hilgemeier, A. Krauss, A. Redder, C. Rolfs, H. P. Trautvetter, and T. R. Donoghue, Nucl. Phys. A415, 93 (1984).
\bibitem{JH98}J. Jos\'e and M. Hernanz, Astrophys. J. 494, 680 (1998). 
\bibitem{Cas11}J. Casanova, J. Jos\'e, E. Garc\'ia-Berro, S.N. Shore, and A.C. Calder, Nature 478, 490 (2011).
\bibitem{Kar14} A. Karakas, MNRAS (submitted, 2014).
\bibitem{Can93} R.C. Cannon, MNRAS 263, 817 (1993).


\end{thebibliography}
\end{document}